\documentclass[nofootinbib,floats,aps,superscriptaddress,preprint]{revtex4}
\usepackage{graphicx}
\newcommand{\bea}{\begin{eqnarray}}
\newcommand{\eea}{\end{eqnarray}}
\newcommand{\beqa}{\begin{eqnarray}}
\newcommand{\eeqa}{\end{eqnarray}}
\newcommand{\beq}{\begin{equation}}
\newcommand{\eeq}{\end{equation}}
\newcommand{\bay}{\begin{array}}
\newcommand{\eay}{\end{array}}

\newcommand{\no}{\nonumber}

\newcommand{\ra}{\rangle}
\newcommand{\la}{\langle}

\def\del{\partial}

\def\gsim{\ \rlap{\raise 3pt \hbox{$>$}}{\lower 3pt \hbox{$\sim$}}\ }
\def\lsim{\ \rlap{\raise 3pt \hbox{$<$}}{\lower 3pt \hbox{$\sim$}}\ }

\newcommand{\z}{z}

\begin{document}

\preprint{\vbox 
{\hbox{} 
\hbox{} 
\hbox{} 
\hbox{SLAC-PUB-10552} 
\hbox{LBNL-55326}
\hbox{UCB-PTH-04/20}
\hbox{hep-ph/0407260}}}

\vspace*{3cm}

\title{\boldmath Twisted Split Fermions}

\def\addtech{Department of Physics,
Technion--Israel Institute of Technology,\\
Technion City, 32000 Haifa, Israel\vspace*{6pt}}
\def\addslac{Stanford Linear Accelerator Center, 
Stanford University, Stanford, CA 94309\vspace*{6pt}}
\def\adducsc{Santa Cruz Institute for Particle Physics, 
University of California, Santa Cruz, CA 95064\vspace*{6pt}}
\def\addlbl{Theoretical Physics Group,
Lawrence Berkeley National Laboratory, \\
University of California, Berkeley, CA 94720\vspace*{6pt}}
\def\adducb{Department of Physics, University of California,
Berkeley, CA 94720\vspace*{20pt}}

\author{Yuval Grossman}\affiliation{\addtech}\affiliation{\addslac}
\affiliation{\adducsc}
\author{Roni Harnik}\affiliation{\addlbl}\affiliation{\adducb}
\author{Gilad Perez}\affiliation{\addlbl}
\author{Matthew D. Schwartz}\affiliation{\addlbl}\affiliation{\adducb}
\author{Ze'ev Surujon}\affiliation{\addtech}

\begin{abstract} \vspace*{8pt}
The observed flavor structure of the standard model arises naturally
in ``split fermion'' models which localize fermions at different
places in an extra dimension. It has, until now, been assumed that the
bulk masses for such fermions can be chosen to be flavor diagonal
simultaneously at
every point in the extra dimension, with all the flavor violation
coming from the Yukawa couplings to the Higgs. We consider the more
natural possibility in which the bulk masses cannot be simultaneously
diagonalized, that is, that they are twisted in flavor
space. We show that, in general, this does not disturb the natural
generation of hierarchies in the flavor parameters. Moreover, it is
conceivable that all the flavor mixing and CP-violation in the
standard model may come only from twisting, with the five-dimensional
Yukawa couplings taken to be universal.
\end{abstract}

\maketitle

\section{Introduction}
One of the motivations to extend the Standard Model (SM) is to explain
the fermion flavor structures. It is likely that there is a more
fundamental theory that produces the observed masses and mixing angles
in a natural way, namely, without small dimensionless parameters. One
such framework uses split fermions to generate the small
numbers~\cite{AS}.  The basic idea is to localize the SM fermion
fields at different locations in compact extra dimensions. Then, the
four dimensional (4D) Yukawa couplings between left handed and right
handed 4D fermion fields are exponentially suppressed by the overlap
of the corresponding zero mode wavefunctions. In general, such a split
fermions setup induces small and hierarchical 4D Yukawa couplings
without imposing any extra symmetries. In addition, one can account
for proton stability by separating quarks and leptons in the extra
dimension. Specific realizations, phenomenological implications and
experimental signatures of the split fermions framework can be found in
\cite{AS,MS,GGH,KT,PDG,ASmodels,GT,Mats,Branco,GP,GrPe,GaussLocH,LocH,NP,AGS}.

Fermion localization works as follows \cite{AS,MS,PDG}. Consider, for
simplicity, a model with one infinite extra dimension. (For the more
realistic case of a finite extra dimension see, for example,
\cite{GGH,KT,GrPe}.) The model contains one bulk scalar, the
localizer, which is assumed to get a vacuum expectation value (vev)
which depends on the extra dimension coordinate, $\z$. Thus, the five
dimensional (5D) Dirac spinors have two mass terms: a $\z$-independent 
bare mass and a $\z$-dependent mass term due to the
couplings to the localizer. For $\Psi_i$, a generic Dirac field (with
$i=1,2,3$ as the generation index) and $\Phi(\z)$, the localizer vev,
these mass terms read
\beq   \label{genL}
{\cal L}=\bar\Psi^i \tilde M_{ij}(\z)\Psi^j, \qquad
\tilde M_{ij}(\z)\equiv m_{ij}+\lambda_{ij}\Phi(\z)\,,
\eeq
where $m_{ij}$ and $\lambda_{ij}$ are $\z$-independent Hermitian
matrices. 

In the past is was always assumed that $m_{ij}$ and $\lambda_{ij}$ can be
diagonalized simultaneously. In that case the problem of obtaining the 4D
observables is significantly simpler.  We can carry out the Kaluza-Klein (KK)
decomposition of the 5D fields in the basis in which the mass matrix is
diagonal.  Each zero mode, which is interpreted as the corresponding SM
chiral fermion field, then has a wavefunction in the extra dimension
$f_i(\z)$.
These wavefunctions must satisfy the condition derived from the 5D
Dirac equation:
\beq
\left[\partial_z-\tilde M_i (\z)\right]f_i(\z)=0, \qquad
\tilde M_i = [m_{ii}+\lambda_{ii}\Phi(\z)]
\label{zero}
\,,
\eeq
where $\partial_z \equiv \partial/\partial z$. The solution is
\beq
{f_i}(\z)=f_i(0)\exp\left[\int_0^{\z} \tilde M_i(\z')d\z'\right]\,,
\label{formal}
\eeq
where $f_i(0)$ is an arbitrary vector that specifies the values of the
solutions at $\z=0$.  For example, in the case where the localizer vev
is linear, the zero mode wavefunctions are Gaussians
\cite{AS,MS}. These Gaussians peak at the points where $\tilde
M_i(\z)=0$. The fact that the wavefunctions which correspond to
different generations are localized at different points leads to the
required small overlaps.

In this work we relax the assumption that the parameters in the Lagrangian
should be aligned in flavor space, that is, we assume that $m_{ij}$ and
$\lambda_{ij}$ \emph{cannot} be diagonalized simultaneously.  We call the
unaligned case ``twisted'', and the aligned case ``untwisted''.  We will
focus on answering the following two questions related to the attractive
features of the split fermions framework:
\begin{itemize}
\item[(i)]
In the presence of twisting, can one naturally 
suppress operators which involve 
fields in different representations?
This is required to account for the proton longevity.
\item[(ii)]
In the presence of twisting, are the
hierarchies between operators which involve  
fields from the same representation still natural?
This is required to account for the flavor puzzle.
\end{itemize}
As we demonstrate below the answer to both of the above questions is
positive: The presence of twisting does not spoil the basic
appealing features of the split fermions framework.

Once we understand how twisting affects hierarchies, we can ask 
whether twisting may also be useful for model building.
Again the answer is positive.  One
example is related to the fact that twisting induces new CP violating
sources. This was used in~\cite{NP} to construct a new type of
leptogenesis model.  Below we demonstrate how the standard model
flavor mixing and CP violation may arise purely from twisting.
Another possible application is related to a solution of the strong CP
problem, which will be discussed in a separate work.

\section{The Twist - Basic formalism}\label{Form}

We consider the most general Lagrangian of Eq. (\ref{genL}). We would
like to find the profile of the zero modes and calculate the 4D
observables in terms of the 5D Lagrangian parameters. The answer is
nontrivial because the effective mass matrix of the
fermions $\tilde M_{ij}(\z)$ is a $\z$-dependent Hermitian
matrix. There is no global $SU(3)$ flavor transformation which brings
$\tilde M_{ij}(\z)$ to a diagonal form simultaneously at every
$\z$. $\tilde M_{ij}(\z)$ can be diagonalized by a $\z$-dependent
special unitary matrix $U(\z)$, however this does not leave the 5D kinetic
terms invariant. This is to be compared to the
untwisted case where $U$ is $\z$-independent.
Formally we can introduce a local measure of the twist by
\beq
   R_{ij}(\z)=\left[\tilde M_{ik}(\z),
   {d\over d\z}\,\tilde M_{kj}(\z)\right].
\eeq
When $R_{ij}(\z)$ is non-zero, a twist is
present in that region. Only when $R_{ij}(\z)=0$ for all values of
$\z$ does the general (twisted) case reduce to the flavor-aligned 
(untwisted) case.

In the general twisted case the KK decomposition for a vector (in flavor
space) fermion field is given by
\beq
   \Psi_i(x_\mu,\z)=\sum_{n,\alpha} (f_L^n)_{i\alpha}(\z) P_L \psi^n_\alpha(x_\mu) +
   \sum_{n,\alpha} (f_R^n)_{i\alpha}(\z) P_R \psi^n_{\alpha}(x_\mu) \,,
\eeq
where $x_{\mu}$ are the known four dimensions, $(f_L^n)_{i\alpha}$ and
$(f_R^n)_{i\alpha}$ are $\z$-dependent wavefunctions, and
$P_L\,(P_R)$ is the left (right) handed chirality projection operator.
Our notation is such that 5D (4D) fields are denoted by
capital (lowercase) letters. We can think about the Latin (Greek)
indices as labeling the flavor space in 5D (4D). 
Note that 
the 5D wavefunctions, $f_R^n$ and
$f_L^n$, are generic $\z$-dependent matrices in flavor space, as 
compared to the untwisted case \cite{AS}, where they were diagonal
and hence simply functions.

We are interested in the zero modes since they correspond to the SM fermions.
Consider, for example, the wavefunctions of the left handed zero
modes, $f_{j\alpha}(\z)$.
These must satisfy the condition derived from the 5D Dirac equation:
\beq
   \left[\delta_{ij}\partial_z-\tilde M_{ij}(\z)\right]f_{j\alpha}(\z)=0
   \label{zero-gen}  \,,
\eeq
where
$i,j=1\ldots 3$ stand for the three components of a single wave
function, while $\alpha=1\ldots 3$ labels the three
different solutions to this equation and thus corresponds to the SM flavor
indices.
The solution may be written formally in a straightforward way:
\beq
   f_{i\alpha}(\z)= P \exp\left[\int_0^{\z} \tilde
   M_{ij}(\z')d\z'\right] \times f_{j\alpha}(0)\,,
   \label{formal-twist}
\eeq
where $P$ stands for the path ordered product and $f_{j\alpha}(0)$ is
an arbitrary $3\times 3$ matrix that specifies the values of the
different solutions at $\z=0$. In principle, we can choose $f_{j\alpha}(0)$
such that the three vectors $f_{i1}$, $f_{i2}$
and $f_{i3}$ constitute a set of orthonormal eigenvectors:
\beq
   \int d \z \,f^*_{i\alpha}(\z)\,f_{i\beta}(\z)=\delta_{\alpha\beta}\,.
\eeq
In practice, we can just choose the vectors to be linearly
independent and then get an orthonormal set by applying the
Gram-Schmidt procedure.

The nontrivial difference between 
the twisted (\ref{formal-twist}) and untwisted (\ref{formal}) case
is that with twisting 
one cannot factorize the solution of the zero modes into a form of
flavor-space times $\z$-space. 
At each point, each solution is a 
vector in flavor space, but the twisting forces it to
rotate (twist) as it moves along the extra-dimension.
This follows directly from the fact that with twisting
 the matrix $\tilde M_{ij}$
cannot be diagonalized simultaneously at all $\z$.

Unfortunately, in the most general twisted case,
equation (\ref{formal-twist}) is not very instructive (although
it can be used for numerical computations).
In certain cases even with twisting we can find explicit solutions. 
For example, in two generations,
the two coupled first-order differential equations (\ref{zero-gen}) 
can be combined into a single
second order equation. 
Then, if $\tilde M$ has a simple enough form, we may be able to find 
solutions. For example, if $\tilde M$ depends linearly on $\z$, 
the solutions are Kummer functions (see Appendix \ref{ap}).
In three generations the composite equation is third order and
is in general unsolvable, even with a linear $\tilde M$.

In order to obtain the 4D observables we start from the 5D couplings of
the fermions to the SM Higgs field. Consider for example the couplings
of the quarks doublets, $Q_i$, to the down type quark singlets, $D_j$,
\beq
   S=\int d^5 x Y^{d}_{ij} H \bar Q_i D_j\,,
\eeq
where $H$ is the SM Higgs field
and the Dirac structure is suppressed. The 5D Yukawa
couplings, $Y^{d}_{ij}$, are assumed to be arbitrary parameters
without any specific flavor structure.  Performing the KK reduction,
assuming that the profile of the Higgs vev is flat, and keeping only
the zero modes, we obtain the standard 4D action
\beq
   S=\int d^4 x\,y^d_{\alpha\beta}\, h \, \bar q_\alpha \, d_\beta\,.
\eeq
The dimensionless 4D Yukawa couplings are given by
\beq
   \label{4Dyukawa}
   y^d_{\alpha\beta}= \int d \z Y^{d}_{ij} \,f^{q*}_{i\alpha}(\z)\,
   f^d_{j\beta}(\z)\,,
\eeq
where $f^q$ ($f^d$) is the wavefunction of the left (right) handed
quark doublet (down type singlet) and the sum over $i$ and $j$ is
implicit. 

For simplicity, here and in it what follows, we work with rescaled
parameters. That is, we scale constants and wavefunctions to the
appropriate power of the fundamental scale to make them dimensionless.


\section{General properties of the zero modes}\label{ana}

In the following two sections we demonstrate that twisting does not
destroy the essential appealing features of the split fermions
framework, namely, localization and separation of the fermion zero
modes.  

First, consider localization. We confine our discussion to the case
were the eigenvalues of $\tilde M_{ij}(\z)$ are monotonically
decreasing functions of $\z$ (as is the case in the models of
\cite{AS}). In the untwisted case the peaks of the zero modes are
located at the points where one of the flavors has a zero bulk
mass. This implies that we can separate fermions in different
representations, for example quarks and leptons, to forbid proton
decay. In this section we show that a similar localization holds in the
twisted case as well.

To see this, we show that since the eigenvalues of $\tilde M_{ij}(\z)$
are monotonic functions of $\z$, we can define a ``localization
region''.  We define this region as the region between the two points
$z_a$ and $z_b$ such that for $z<z_a$ ($z>z_b$) all the eigenvalues
are positive (negative). The magnitude of each of the zero modes,
$|f_\alpha|^2\equiv \sum_i f^*_{i\alpha} f_{i\alpha}$, decays outside
the localization region as is shown below.

It is sufficient to study the norm of the wavefunction because the overlap
between the norms provides an upper bound on the overlap between the actual
wavefunctions,
\begin{equation}
   \label{ineq}
   \int d\z f_{i\alpha}^{q*}(\z) f_{j\beta}^u(\z) \le
   \int d\z |f_{\alpha}^{q}(\z)| |f_{\beta}^u(\z)|,
\end{equation}
which serves as a bound on the effective 4D couplings. Using
eq. (\ref{zero-gen}) we obtain
\begin{equation}
\label{norm}
\del_5 |f_\alpha(\z)|^2= 
2\sum_k m_k(\z) |f_{k\alpha}(\z)|^2.
\end{equation}
where there is no sum on
the index $\alpha$ and $m_k(\z)$ are the eigenvalues of $\tilde
M_{ij}(\z)$. We denote by
$m_{\max}(\z)$ [$m_{\min}(\z)$] the maximal [minimal] eigenvalue of
$\tilde M$ at $\z$. Then, we can place bounds on the right hand side of
equation~(\ref{norm}) 
\begin{equation}
m_{\min} |f_\alpha(\z)|^2 \le \sum_i m_i(\z) |f_{i\alpha}(\z)|^2 \le 
m_{\max} |f_\alpha(\z)|^2\,.
\end{equation}
We see that the norms of the zero modes fall at large
positive and negative values. In particular,
\begin{equation}
\label{bound>}
|f_\alpha(\z)|^2\le \exp\left[2{\int_{z_b}^{\z} d\z'
m_{\max}(\z')}\right] |f_\alpha(z_b)|^2 \qquad {\rm for}~\z>z_b.
\end{equation}
Note that $m_{\max}<0$ over all of the integration range.  A similar
bound may be placed in the $\z<\z_a$ region
\begin{equation}
\label{bound<}
|f_\alpha(\z)|^2\le \exp\left[-2\int_{\z}^{z_a} d \z' m_{\min}(\z')\right]
|f_\alpha(z_a)|^2\qquad {\rm for}~\z<z_a,
\end{equation}
where $m_{\min}>0$ all over the integration range.
We see that the norm of the wave function indeed decays outside the
localization region.

The above arguments hold separately for each SM representation.  
Thus, different representations can be separated if their localization
regions do not overlap.
In models where the localizer vev is linear, like that of~\cite{AS},
$m_{\max}$ and $m_{\min}$ are roughly linear far away from the
localization region.  In that case the bounds in equations
(\ref{bound>}) and (\ref{bound<}) imply that far from the localization
region the norms of the zero modes are suppressed exponentially (as
Gaussians).
The generalization to the more realistic models in which a stable
scalar configuration is not a monotonic function of $\z$ is
straightforward.  In these cases one can divide the extra dimension
into regions where the scalar is monotonic and apply the above
analysis for each of the regions separately.  Thus we have answered
question (i) from the introduction: fermions in different
representations can be naturally separated.

\section{The Adiabatic Approximation}\label{adi}

Though we have shown that exponentially small overlaps are easily
achieved between different representations, in order to solve the
flavor puzzle hierarchies within a representation are required.  To
see whether this is the case in our framework one needs a better
handle on the profiles of fermions zero modes.  As we already
mentioned, the general solution to the zero mode wavefunction equation
is not very useful in this respect. Here we show that in many cases
one can make an approximation in which the governing physics is clear.

The zero mode equation (\ref{zero-gen}) bears resemblance to the time
dependent Schr\"odinger equation, $i\del_t |\psi\rangle = H
|\psi\rangle$, with $it$ replaced by $\z$. The $\z$ dependent $\tilde
M$ corresponds to a time varying Hamiltonian.  One of the useful
approximation methods for solving a time varying Hamiltonian in
quantum mechanics is the adiabatic approximation \cite{messiah} where
the wavefunction is assumed to be an instantaneous energy eigenstate
at all times. 

It is useful to make a similar approximation in our case.  The
assumption we make is that the independent solutions of eq.
(\ref{zero-gen}) are each governed by a single eigenvalue of the
mass matrix at each point in the extra dimension. We thus expect each
solution to be aligned with one of the eigenvectors of $\tilde M(\z)$
at each point.

We begin by writing an ansatz for the solution to the equation which
satisfies the guideline mentioned above. The ansatz for the left
handed zero mode profile is
\beq
f^{\mathrm{ad}}_{i\alpha}\left(\z\right)=
{1\over N}\,\exp\left[\int_0^{\z} d\z'\,
  m_\alpha\left(\z'\right)\right]V_{i\alpha}
\left(\z\right) 
\label{adisol-fi}\,,
\eeq
where $V_{i\alpha}(\z)$ are the $\z$ dependent normalized
eigenvectors of the twisted bulk mass matrix
\beq
\tilde M_{ij}(\z) V_{j\alpha}\left(\z\right)=
m_\alpha (\z) V_{i\alpha} (\z) \qquad \mbox{with} \qquad 
V_{i\alpha}^* V_{i\beta}=\delta_{\alpha\beta}\,. 
\eeq
Within this approximation each zero mode is localized around
the zero of a single ($\z$ dependent) eigenvalue of the mass matrix.
The adiabatic ansatz is thus a straightforward generalization of the
untwisted solution. In both cases the wavefunction of each profile is
governed by a single function which is an eigenvalue of the bulk mass
matrix and therefore in both cases the wavefunctions are localized.
In the adiabatic limit the only added feature
introduced with twisting is that the wavefunction points in different
directions in flavor space for various values of $\z$.

To what extent is the adiabatic approximation valid in generic models?  
The standard condition in quantum mechanics literature is that the
approximation holds so long as the quantity
\begin{equation}
\label{non-adiab}
\frac{\langle V_1|\del_z \tilde M|V_2\rangle}{(m_1-m_2)^2},
\end{equation}
is small throughout the evolution of the system~\cite{messiah}. 
Our case however, is somewhat more subtle due to the fact that the
evolution of the wavefunctions with `time' is not unitary. The rate at
which the true solution deviates from the adiabatic one at a certain point is
indeed proportional to the quantity in eq.~(\ref{non-adiab}) but is
also depends on the values of the wavefunctions at that point.

\begin{figure}[t]
\centerline{\includegraphics[width=.5\textwidth]{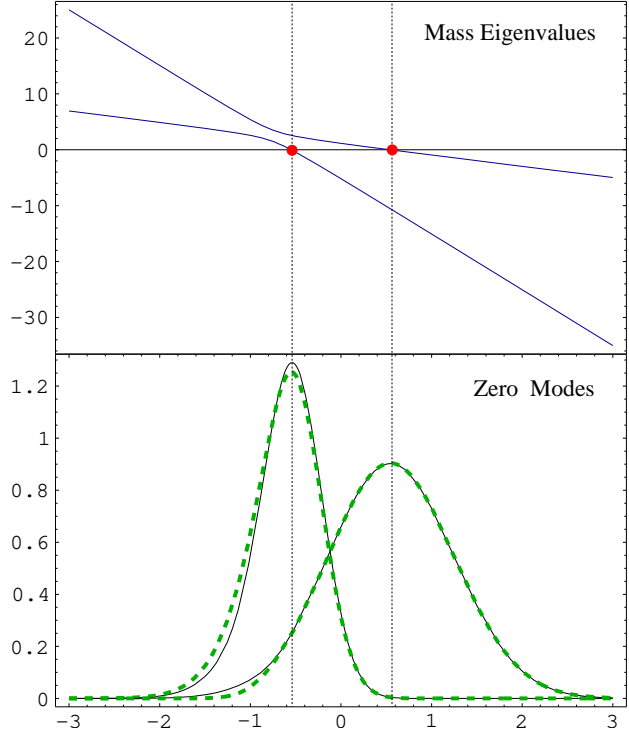}}
\caption{ The two solid blue lines in the upper figure are the eigenvalues of
the mass matrix $\tilde M(\z)$ given in eq.~(\ref{figmass}). The
localization region defined in the previous section lies between the two
vertical dotted lines.  The exact solution to the norm of the
wavefunctions is in the
lower figure in solid lines while the adiabatic approximation is in thick
dashed green curves. Note that the departure of the adiabatic solution from
the exact one occurs where the two eigenvalues approach one another.} 
\label{figm12} 
\end{figure}

We did not try here to fully formulate and derive the necessary
conditions for the validity of the above approximation. One
can, however, get good intuition by numerically comparing it with the exact
solution, eq.  (\ref{formal-twist}).  This allows us to learn about
the accuracy of the adiabatic approximation.  As an example consider
the following mass matrix for the two generation case
\beq \label{figmass}
\tilde M(\z) =\pmatrix{-2z+1  & 1 \cr 1 & -5z-10}\,,
\eeq
where we work with rescaled parameters such that $\z$ and $\tilde M$
are dimensionless.  In the upper part of figure~\ref{figm12} we plot the
two eigenvalues of the mass matrix and in the lower part of the
figure~\ref{figm12} we plot the norm of adiabatic solution that is
induced by the eigenvalues (green dashed curves). For comparison the
adiabatic curves are plotted on top of the exact solution (black solid
lines).

We see that in the case presented in figure~\ref{figm12} the adiabatic
approximation is very good. It noticeably deviates from the exact
solution only in the region where the two eigenvalues approach one
another as one would expect from eq.~(\ref{non-adiab}). This departure
only occurs once both profiles have decayed (or have started to do
so). Consequently, physical quantities are hardly affected and the
approximation holds to a very good accuracy.

We expect this behavior to be generic in models that have well localized and
separated wavefunctions. This is because separation implies that at the
points $z_i$, where any of the mass eigenvalues vanishes, the mass
differences are large.  Thus, non-adiabatic contributions are supressed near
$z_i$. If the eigenvalues are changing smoothly the non-adiabaticity
generically occurs well outside the localization region.  We have indeed
observed that increasing the separation (and hierarchy) between flavors
improves the approximation.  We have checked numerically several cases that
lead to hierarchical Yukawa matrices and found that the adiabatic
approximation works very well in these cases.

\section{Hierarchies and mixing from twisting}

So far we have shown that even with twisting, localization and
separation within a representation naturally occur.  In this section
we discuss twisting as the only source of flavor mixing and CP
violation. We consider below the following two questions:
\begin{itemize}
\item[(a)]
Can all the flavor mixing and CP violation come from twisting?
That is, can we take the 5D Yukawa matrices to be universal, so that
all of the SM flavor structure comes
from twisting? 
\item[(b)] Can the above scenario be realized naturally?
\end{itemize}

We show that the answer to the above two questions is positive.  We
first demonstrate numerically that twisting can serve as the only
source of flavor mixing. Then we support this observation, for more
generic cases, using the adiabatic approximation.  Finally, we present
a toy model in which this situation occurs naturally. 

Let us begin with the numerical example. We denote the
mass matrices as
\beq\label{efM-two-we}
\tilde M_{ij}^r\left(\z\right)=m_{ij}^r- \eta_{ij}^r \z
\,.
\eeq
with $r=Q,U,D$.
We assume that the 5D Yukawa couplings are
proportional to the unit matrix.
Then, the Lagrangian of the model is
given schematically by
\beq \label{5Dsol}
{\cal L}= Y^U\delta_{ij} H \bar Q_i\, U_j+
Y^D\delta_{ij} H \bar Q_i\, D_j+
\tilde M_{ij}^{Q}\bar Q_i Q_j +
\tilde M_{ij}^{U}\bar U_i U_j+ \tilde
M_{ij}^{D}\bar D_i D_j\,.
\eeq
When $m^r_{ij}$ and $\eta^r_{ij}$ can be diagonalized simultaneously
this model is untwisted.  In the most generic case, however, when no
symmetry is imposed, the mass matrix is expected to be twisted.
For our explicit example we choose the following parameters for the
bulk masses
\beqa
\tilde M^Q(\z)&=&\pmatrix{3.4 \,\z -4& 2.3 \cr 2.3 & 2.6 \,\z - 6}\,,
\no \\[9pt]
\tilde M^D(\z)&=&\pmatrix{4.1 \,\z +1& 1 \cr 1 & 3.1 \,\z - 3}\,,
\no \\[9pt]
\tilde M^U(\z)&=&\pmatrix{4.5 \,\z +4& 1 \cr 1 & 2.9 \,\z + 2}\,,
\label{Matrices}
\eeqa
where we work with rescaled parameters such that $\z$ and $\tilde
M^r$ are dimensionless.  We obtain hierarchical fermion masses
\beq
{m_d \over m_s} = 2.1 \times 10^{-1}, \qquad
{m_u \over m_c} = 8.0 \times 10^{-3},
%
\eeq
and finite mixing
\beq
\sin\theta_C=0.19.
\eeq
Note that this is only an example. We did not try to search for a mass
matrix which precisely generates the observed 4D flavor parameters.
We use this example only to demonstrate the fact that 4D flavor mixing
and hierarchical masses can originate only from the twisting.

While we gave an explicit example only for one simple case, the
conclusion holds much more generally. To see this, we use the adiabatic
approximation. Using Eqs. (\ref{4Dyukawa}) and (\ref{adisol-fi}) 
we see that 
\beq \label{generic-xx}
y^d_{\alpha\beta} \approx {1\over N_Q N_D} \int d \z
\exp\left[\int_0^{\z} d\z'\,
  m^Q_\alpha\left(\z'\right)\right]
\exp\left[\int_0^{\z} d\z'\,
  m^D_\beta\left(\z'\right)\right]
V_{i\alpha}^Q\left(\z\right) V_{i\beta}^{D*}\left(\z\right) \,.
\eeq
The exponential factors in (\ref{generic-xx}) teach us that we expect
a hierarchical structure for the 4D Yukawa couplings. In addition, for
the realistic case of three generations, the product ${V^Q_{i\alpha}}
V^{D*}_{i\beta}$ is just a product of two $SU(3)$ matrices.  This
shows that flavor mixing is induced.  Furthermore, these $SU(3)$
matrices generically contain ${\cal O}$(1) phases. Thus, $\delta_{\rm
CKM}$, the CP violating phase in the 4D CKM matrix, is expected to be
of order unity as observed.

The above numerical study exemplifies the possibility
that all of the SM flavor conversion is achieved due to the twist in
the bulk, and not due to the 5D Yukawa couplings.  Below we construct
a toy model which naturally realizes this idea.
Consider a model for quarks with a non-Abelian horizontal flavor
symmetry $SU(3)_F$ on a 5D orbifold $M_4\times S_1/Z_2$.  The
fermions, $Q^i,U^i,D^i$ are fundamentals of the flavor group where all
other SM fields are singlets. The Higgs field, the left handed
component of $Q^i$ and the right handed component of $D^i$ and $U^i$
are even under the $Z_2$ orbifold symmetry, and all the other fields
are odd.  Because of the orbifold symmetry, a bulk mass for the
fermions is forbidden.  But an effective mass can be generated from
the vev of $Z_2$ odd bulk scalars that are SM singlets. For our
example, we include an adjoint (octet) $\Phi_{ij}$ and a singlet
$\phi$ of $SU(3)_F$.  The Yukawa couplings to the Higgs, on the other
hand, are allowed and are proportional to unit matrices due to the
flavor symmetry.

Generically, there is a potential for the scalars in the bulk. This
potential naively generates an untwisted scalar vev since varying vevs
are usually not the lowest energy configuration. Thus, we also include
boundary terms that change this naive expectation.  When the symmetry
is explicitly broken on the boundaries, the competition between the
bulk and boundary terms can force the vev of $\Phi$ to be twisted.

To be explicit, consider the case in which the boundaries preserve
only an $SU(2)$ subgroup of the bulk $SU(3)_F$. Under this $SU(2)$,
$\Phi_{ij}$ decomposes into a scalar $\Phi\equiv\Phi_{33}$, a
fundamental $\Phi_l=\left( \Phi_{13}, \Phi_{23}\right)$, an
antifundamental $\Phi_l^*$, and an adjoint $\Phi_{lm}$ with $l,m=1,2$.
For each of the above fields we can write a boundary term
\beq
-{\cal L}^{\rm brane}_\Phi=\sum_{B=0,\pi R}\left[ a_{B}^2\left(\partial_z
\Phi\right)^2 +b_{B}^2\left(\partial_z
\Phi_l\right)^2+c_{B}^2\left(\partial_z \Phi_{lm}\right)^2\right]+\ldots
\eeq
where higher order (stabilizing) terms are omitted for simplicity.
The parameters $a_{0,\pi R}$, $b_{0,\pi R}$ and $c_{0,\pi R}$ are real and
generically different on the two different branes.
The terms above cause non-zero vevs to be developed for the derivative
of $\Phi,\Phi_1$ and $\Phi_{lm}\,$ at $z=0$
\beqa
\langle \partial_z\Phi\rangle|_0={\cal O}(1)
\,,\ \ \ 
\langle\partial_z \Phi_l\rangle|_0=\pmatrix{0 \cr {\cal O}(1)}
\,,\ \ \
\langle\partial_z \Phi_{lm}\rangle|_0=H_{lm}
\,,\label{vevs}
\eeqa
where $H_{lm}$ is an order one Hermitian matrix and we used an $SU(2)$
transformation to bring the vev of $\partial_z\Phi_l$ to its special
form.  Similarly, at $z=\pi R$ the derivatives get non-vanishing vevs,
which are generically different from those at $z=0$. Consequently,
the mass matrix at both boundaries are not aligned and a twist is
generated in the bulk.

We assume that $v^2_F/\Lambda^2 \ll 1$ where $v_F$ is the typical vev
of the bulk scalars and $\Lambda$ is the effective cutoff of the 5D
theory.  Then, to leading order in $v^2_F/\Lambda^2$, the 5D Lagrangian
is given by eq. (\ref{5Dsol}) with
\beq
   \tilde M_{ij}^r=c_1^r \la 
   \Phi_{ij}\ra+c_2^r \la \phi\ra\,,
\eeq
where $c_1^r$ and $c_2^r$ are unknown constants. We see that the 5D
Yukawa couplings are proportional to the unit matrix and that $\tilde
M_{ij}^r$ cannot be globally diagonalized at each $\z$ because it is
twisted.  Due to the fact that the bulk scalars are odd under the
$Z_2$ orbifold symmetry the correction to the universal Yukawa
matrices is suppressed by ${\cal O}(v^2_F/\Lambda^2)$.  There are
additional higher dimension brane Yukawa terms which are volume
suppressed and are therefore negligible.  We conclude that in this toy
model the 4D flavor violation is, to leading order, only due to
twisting.

\section{Discussions and conclusions}\label{Con}

We found that in many ways neglecting the twist is a simplifying
assumption. Twisting does not change the fact that split fermions
generically produce hierarchical Yukawa couplings. At first neglecting
the twist seems unjustified: it changes the symmetry breaking pattern
of the theory.  For example, suppose there is a single fermion
representation in the bulk, with 3 flavors.  When $\tilde
M(\z)\propto{\mathbf 1}_3$ there is an enhanced $U(3)$ flavor
symmetry in the theory. Including a diagonal $\tilde M(\z)$ breaks the
$U(3)$ symmetry down to $U(1)^3$. Including a generic twisted $\tilde
M(\z)$ breaks it further to $U(1)$~\cite{NP}. In untwisted models this
last step occurs only due to the standard model Yukawa interactions.

In the untwisted case the fact that we get hierarchical Yukawa
couplings and small mixing can be understood from symmetry
considerations as follows. The $U(1)^3 \to U(1)$ breaking is due to
non-diagonal 5D Yukawa couplings, which connect fields of different SM
representations, and therefore, the last stage of symmetry breaking
occurs between objects that are separated in the extra
dimension. Thus, the 4D Yukawa couplings are small since they are
suppressed by the small overlap of the separated zero modes.  That is,
the $U(1)^3$ remains as an approximate symmetry, and it is restored
once the zero mode wavefunctions are far away from each other.

In the twisted case the fact that an approximate $U(1)^3$ symmetry is
obtained in the low energy effective theory is less
obvious. Nevertheless, we claim that this remain the case due to the
same reason, namely, symmetry breaking occurs between objects that are
separated in the extra dimension.  In other words, even with the
twist, we have shown that the zero modes are localized and
separated. This implies that the 4D Yukawa couplings are small since
they are proportional to the small overlap of the different zero
modes. Just like in the untwisted case, when the separation is very
large, the $U(1)^3$ symmetry is restored. Thus, the effect of the
twist is only to add new sources of flavor mixing and CP violation.

In conclusion, the main result from our study is that the twist does
not affect the general appealing features of the split fermions
framework, that is, the possibility of naturally creating hierarchies
without symmetries.  Furthermore, it opens a possibility in which the
observed CKM mixing and CP violation arise from twisting and not from
the 5D Yukawa couplings.

\acknowledgments
We thank  Zacharia  Chacko, Walter Goldberger, Lawrence Hall, Markus
Luty, Yasunori Nomura and Natalia Shuhmaher for helpful discussions.
The work of YG is supported in part by a grant from the G.I.F., the
German--Israeli Foundation for Scientific Research and Development, by
the United States--Israel Binational Science Foundation through grant
No.~2000133, by the Israel Science Foundation under grant No.~237/01,
by the Department of Energy, contract DE-AC03-76SF00515 and by the
Department of Energy under grant No.~DE-FG03-92ER40689.
RH was supported in part by the DOE under contract DE-AC03-76SF00098 and in
part by NSF grant PHY-0098840.
GP is supported by the Director, Office of Science, Office of High
Energy and Nuclear Physics of the US Department of Energy under
contract DE-AC0376SF00098.
YG and GP thanks the Aspen center for physics in which part of
this work was done.

\appendix


\section{Linear potential: Explicit solution}
\label{ap}

There is one case where we could find an explicit analytic solution to
the zero mode wavefunctions. This is the case of a two generation
model with an infinite extra dimension and a localizer with a linear
vev.  Here we only sketch the derivation \cite{GS}, and discuss some
of the properties of the solution.
 
In the case under study the mass matrix defined in
(\ref{genL}) can be written as
\beq\label{efM-two}
\tilde M_{ij}(\z)=m_{ij}- \eta_{ij} \z
\,.
\eeq
It is convenient to chose a basis in which $m_{ij}$ is real,
$\eta_{ij}$ is diagonal and $\eta_{11} > \eta_{22}$. Then,
eq. (\ref{zero-gen}) is given explicitly by
\beqa \label{masterone}
f_1'(\z)+(\eta_{11} \z -m_{11}) f_1(\z) - m_{12} f_2(\z) & = & 0 , \\
f_2'(\z)+(\eta_{22} \z -m_{22}) f_2(\z) - m_{12} f_1(\z) & = & 0
,  \label{mastertwo}
\eeqa
where the prime denotes a derivative with respect to $\z$ and the index
$\alpha$ was dropped.
The solution is 
\beqa \label{sol-per}
f_1(\z)&=&p(\z)\left[c_1 m_{12} t(\z) M(a,3/2,w(\z))+c_2
M(a-1/2,1/2,w(\z))\right],\no \\
f_2(\z)&=&p(\z)\left[c_2 m_{12}
t(\z) M(a+1/2,3/2,w(\z))+c_1 M(a,1/2,w(\z))\right],
\eeqa
where $M$ is the Kummer function and
\beqa \label{defforM}
p(\z)&=&\exp\left[-{(m_{11}-\eta_{11} \,\z)^2 \over 2 \eta_{11}}\right],
           \no \\[8pt]
a&=& \frac{\eta_{11}-\eta_{22}+m_{12}^2}{2(\eta_{11}-\eta_{22})}, \no \\[8pt]
t(\z)&=& \z-\frac{m_{11}-m_{22}}{\eta_{11}-\eta_{22}}, \no \\[8pt]
w(\z)&=& {\eta_{11}-\eta_{22} \over 2}\,t(\z)^2.
\eeqa
The normalization condition is
\beq
\int \left[f_1(\z)^2+f_2(\z)^2\right]d \z=1.
\eeq
The set of equations (\ref{masterone}) and (\ref{mastertwo}) has two
independent solutions related to the two independent constants $c_1$
and $c_2$ that appear in (\ref{sol-per}). This degree of freedom in
the solution corresponds to the index $\alpha$ of $f_{i\alpha}$. One
can choose the integration constants such that the two wavefunctions
are orthogonal
\beq
\int d\z \sum_{i=1}^2 f_{i1}(\z) f_{i2}(\z) = 0.
\eeq
In practice we ensure the orthogonality by applying the Gram-Schmidt
procedure on a pair of non-orthogonal wavefunctions.

Arriving to the solution in (\ref{sol-per}) is straightforward. We first use
(\ref{masterone}) and plug it into (\ref{mastertwo}) to arrive at a
second order differential equation for $f_1$. Using $f_1(\z)=g(\z) p(\z)
t(\z)$ this equation is
\beq \label{kummer}
w\,{d^2 g\over dw^2}+(3/2-w) {dg \over dw}- a\,g=0,
\eeq
which is the Kummer equation \cite{book}.
The general solution to (\ref{kummer}) is
\beq
g(w)=g_a\,M(a,3/2,w)+g_b\,{M(a-1/2,1/2,w) \over \sqrt{w}}\,,
\eeq
where $g_a$ and $g_b$ are two independent constants.  To get the solution
for $f_2$ we used the following properties of $M$ \cite{book},
\beqa \label{relMa}
{dM(\alpha,\gamma,w) \over dw}&=&{\alpha\over
\gamma}M(\alpha+1,\gamma+1,w) ,\\ \label{relMb}
{w \over \gamma} M(\alpha+1,\gamma+1,w) &=&
M(\alpha+1,\gamma,w)- M(\alpha,\gamma,w), \\ \label{relMc}
\alpha M(\alpha+1,\gamma+1,w)&=&(\alpha-\gamma)M(\alpha,\gamma+1,w)+
\gamma M(\alpha,\gamma,w).
\eeqa

One can check that in the $m_{12}\to 0$ limit the solution of the
twisted case (\ref{sol-per}) reduces to the solution of the untwisted
case. In that limit $a
\to 1/2$ and we recall the following properties of the Kummer function
\beq
M(0,c,b)=1, \qquad
M(c,c,b)=e^b,
\eeq
for arbitrary $c$ and $b$. Then we get
\beq
f_1(\z)=c_2 \exp\left[-{(m_{11}-\eta_{11} \,\z)^2 \over 2
\eta_{11}}\right],
\qquad
f_2(\z)=c_1 \exp\left[-{(m_{22}-\eta_{22} \,\z)^2 \over 2 \eta_{22}}\right].
\eeq
Taking the two independent solution to be those where either $c_1$ or
$c_2$ vanish, we get the two Gaussian solutions of the untwisted case
\cite{AS}.



\begin{thebibliography}{99}

\bibitem{AS}
N.~Arkani-Hamed and M.~Schmaltz,
Phys.\ Rev.\ D {\bf 61}, 033005 (2000) [hep-ph/9903417].




\bibitem{MS}
E.~A.~Mirabelli and M.~Schmaltz,
Phys.\ Rev.\ D {\bf 61}, 113011 (2000)
[hep-ph/9912265].

\bibitem{GGH}
H.~Georgi, A.~K.~Grant and G.~Hailu,
Phys.\ Rev.\ D {\bf 63}, 064027 (2001)
[hep-ph/0007350].

\bibitem{KT}
D.~E.~Kaplan and T.~M.~Tait,
JHEP {\bf 0111}, 051 (2001)
[hep-ph/0110126].

\bibitem{GT}
B.~Grzadkowski and M.~Toharia,
extra
hep-ph/0401108.



\bibitem{GP}
G.~Perez,
Phys.\ Rev.\ D {\bf 67}, 013009 (2003)
[hep-ph/0208102].


\bibitem{GrPe}
Y.~Grossman and G.~Perez,
Phys.\ Rev.\ D {\bf 67}, 015011 (2003)
[hep-ph/0210053].

\bibitem{NP}
Y.~Nagatani and G.~Perez,
hep-ph/0401070.

\bibitem{PDG}
J.~Hewett and J.~March-Russell,
Phys.\ Rev.\ D {\bf 66}, 010001 (2002).

\bibitem{ASmodels}
H.~V.~Klapdor-Kleingrothaus and U.~Sarkar,
Phys.\ Lett.\ B {\bf 541}, 332 (2002)
[hep-ph/0201226];
P.~Q.~Hung and M.~Seco,
Nucl.\ Phys.\ B {\bf 653}, 123 (2003)
[arXiv:hep-ph/0111013];
Y.~Uehara,
JHEP {\bf 0112}, 034 (2001)
[hep-ph/0107297];
G.~Barenboim, G.~C.~Branco, A.~de Gouvea and M.~N.~Rebelo,
Phys.\ Rev.\ D {\bf 64}, 073005 (2001)
[hep-ph/0104312];
J.~L.~Crooks, J.~O.~Dunn and P.~H.~Frampton,
Astrophys.\ J.\ {\bf 546}, L1 (2001)
[astro-ph/0002089];
%
S.~Khalil and R.~Mohapatra,
hep-ph/0402225;
P.~Q.~Hung, M.~Seco and A.~Soddu,
Nucl.\ Phys.\ B {\bf 692}, 83 (2004)
[arXiv:hep-ph/0311198];
J.~M.~Frere, G.~Moreau and E.~Nezri,
along
Phys.\ Rev.\ D {\bf 69}, 033003 (2004)
[hep-ph/0309218];
B.~Lillie and J.~L.~Hewett,
Phys.\ Rev.\ D {\bf 68}, 116002 (2003)
[hep-ph/0306193];
W.~F.~Chang and J.~N.~Ng,
JHEP {\bf 0212}, 077 (2002)
[hep-ph/0210414];
P.~Q.~Hung,
Phys.\ Rev.\ D {\bf 67}, 095011 (2003)
[hep-ph/0210131].

\bibitem{Mats}
T.~Matsuda,
Phys.\ Rev.\ D {\bf 66}, 047301 (2002)
[hep-ph/0205331];
Phys.\ Rev.\ D {\bf 66}, 023508 (2002)
[hep-ph/0204307];
Phys.\ Rev.\ D {\bf 65}, 107302 (2002)
[hep-ph/0202258].



\bibitem{GaussLocH}
N.~Haba and N.~Maru,
Phys.\ Rev.\ D {\bf 66}, 055005 (2002)
[arXiv:hep-ph/0204069];
Mod.\ Phys.\ Lett.\ A {\bf 17}, 2341 (2002)
[arXiv:hep-ph/0202196];
N.~Haba, N.~Maru and N.~Nakamura,
hep-ph/0209009;
N.~Maru,
Phys.\ Lett.\ B {\bf 522}, 117 (2001)
[hep-ph/0108002];
M.~Kakizaki and M.~Yamaguchi,
Int.\ J.\ Mod.\ Phys.\ A {\bf 19}, 1715 (2004)
[arXiv:hep-ph/0110266];
M.~Maru, N.~Sakai, Y.~Sakamura and R.~Sugisaka,
Nucl.\ Phys.\ B {\bf 616}, 47 (2001)
[hep-th/0107204];
M.~Kakizaki and M.~Yamaguchi,
Prog.\ Theor.\ Phys.\ {\bf 107}, 433 (2002)
[hep-ph/0104103].


\bibitem{LocH}
D.~E.~Kaplan and T.~M.~Tait,
JHEP {\bf 0006}, 020 (2000)
[hep-ph/0004200];
G.~R.~Dvali and M.~A.~Shifman,
Phys.\ Lett.\ B {\bf 475}, 295 (2000)
[hep-ph/0001072];
A.~Hebecker and J.~March-Russell,
Phys.\ Lett.\ B {\bf 541}, 338 (2002)
[hep-ph/0205143];
F.~Del Aguila and J.~Santiago,
JHEP {\bf 0203}, 010 (2002)
[hep-ph/0111047].

\bibitem{Branco}
G.~C.~Branco, A.~de Gouvea and M.~N.~Rebelo,
Phys.\ Lett.\ B {\bf 506}, 115 (2001)
[hep-ph/0012289].






\bibitem{AGS}
N.~Arkani-Hamed, Y.~Grossman and M.~Schmaltz,
Phys.\ Rev.\ D {\bf 61}, 115004 (2000) [hep-ph/9909411];
T.~Han, G.~D.~Kribs and B.~McElrath,
Phys.\ Rev.\ Lett.\  {\bf 90}, 031601 (2003)
[arXiv:hep-ph/0207003];
W.~F.~Chang, I.~L.~Ho and J.~N.~Ng,
Phys.\ Rev.\ D {\bf 66}, 076004 (2002)
[arXiv:hep-ph/0203212];
S.~Nussinov and R.~Shrock,
Phys.\ Lett.\ B {\bf 526}, 137 (2002)
[hep-ph/0101340];
Phys.\ Rev.\ Lett.\ {\bf 88}, 171601 (2002)
[hep-ph/0112337];
D.~J.~Chung and T.~Dent,
Phys.\ Rev.\ D {\bf 66}, 023501 (2002)
[hep-ph/0112360];
T.~G.~Rizzo,
Phys.\ Rev.\ D {\bf 64}, 015003 (2001)
[hep-ph/0101278];
A.~Masiero, M.~Peloso, L.~Sorbo and R.~Tabbash,
Phys.\ Rev.\ D {\bf 62}, 063515 (2000)
[hep-ph/0003312];
A.~Delgado, A.~Pomarol and M.~Quiros,
JHEP {\bf 0001}, 030 (2000)
[hep-ph/9911252].


\bibitem{messiah}
See for example:  A. Messiah, Quantum Mechanics,Vol. 2, North Holland,
Amsterdam, 1961; L. I. Schiff, Quantum Mechanics, McGraw - Hill, New York
1968.

\bibitem{GS}
N. Shuhmaher, unpublished notes.

\bibitem{book}
See, for example, I.S. Gradshtein and I.M. Ryzhik, fifth edition,
section 9.2.


\end{thebibliography}
\end{document}